%% file: paper.tex
\documentclass{ceurart}

\usepackage{listings}
\usepackage{graphicx}
\usepackage{xcolor}
\usepackage{soul}\setuldepth{article}

\usepackage{acro}
\input{acronyms}

\begin{document}

\copyrightyear{2026}
\copyrightclause{Copyright for this paper by its authors.
  Use permitted under Creative Commons License Attribution 4.0
  International (CC BY 4.0).}

\conference{SAGE 2026: The International Workshop on Semantic Architectures
  for Governance and Explainability, September 15, 2026, Ghent, Belgium}

\title{Does SRL Pave the Road to Explainable Reasoning? Lessons Learned
  from an Implementer's Perspective}

\author[1]{Lander Maes}[%
  orcid=0009-0001-5242-5508,
  email=lander.maes@ugent.be,
]
\cormark[1]
\author[1]{Bryan-Elliott Tam}[%
  orcid=0000-0003-3467-9755,
  email=bryanelliott.tam@ugent.be,
]
\author[1]{Jitse De Smet}[%
  orcid=0009-0002-6513-5013,
  email=jitse.desmet@ugent.be,
]
\author[1]{Jos De Roo}[%
  orcid=0000-0001-8862-0666,
  email=Jos.DeRoo@UGent.be,
]
\author[1]{Pieter Colpaert}[%
  orcid=0000-0001-6917-2167,
  email=pieter.colpaert@ugent.be,
]
\author[1]{Ruben Taelman}[%
  orcid=0000-0001-5118-256X,
  email=ruben.taelman@ugent.be,
]
\address[1]{Ghent University -- imec, Belgium}

\cortext[1]{Corresponding author.}

\input{section/abstract}


\maketitle

\input{section/introduction}

\input{section/related_work}
\input{section/shacl_rules}
\input{section/implementations}
\input{section/findings}

\input{section/conclusion}

\bibliography{references}

\end{document}

%% file: acronyms.tex
\DeclareAcronym{SHACL}{
  short = SHACL,
  long  = Shapes Constraint Language,
}
\DeclareAcronym{RDF}{
  short = RDF,
  long  = Resource Description Framework,
}
\DeclareAcronym{RDFS}{
  short = RDFS,
  long  = RDF Schema,
}
\DeclareAcronym{OWL}{
  short = OWL,
  long  = Web Ontology Language,
}
\DeclareAcronym{SPARQL}{
  short = SPARQL,
  long  = SPARQL Protocol and RDF Query Language,
}
\DeclareAcronym{IRI}{
  short = IRI,
  long  = Internationalized Resource Identifier,
  long-plural-form = Internationalized Resource Identifiers,
}

%% file: section/abstract.tex
\begin{abstract}
    The Shape Rules Language (SRL) Working Draft defines how to derive new RDF triples
    from an RDF graph using inference rules. Each rule matches graph
    patterns and instantiates triple templates whose output feeds into
    validation pipelines, SPARQL queries, or further inference.
    RDF reasoning has traditionally relied on fixed entailment regimes
    (RDFS, OWL), rule-based ad-hoc languages such as N3, or other
    implementation-specific solutions without a shared standard.
    SRL introduces user-defined production rules
    with a defined grammar, dependency analysis, execution ordering, and
    termination guarantees. However, no authoritative implementation exists,
    leaving practitioners with little guidance on how to build a conformant
    engine or on what problems the language can solve.
    We implemented two SRL engines and evaluated both on
    classical RDF reasoning tasks for soundness, completeness, and speed.
    The first reuses an existing SPARQL query engine and its query parser;
    the second is a dedicated engine.
    The SPARQL-based engine reused an existing modular parser for query construction and SPARQL
    CONSTRUCT for triple production, reducing engine-specific work. The dedicated engine was
    two to six times faster, the gap widening as rule sets grow.
    Both engines were validated against the SRL conformance test suite,
    supplemented by additional use-case-driven tests.
    A usable SRL engine can be built inexpensively on top of a
    SPARQL engine, with a moderate speed trade-off that a dedicated
    implementation recovers. Despite the specification's immaturity, the
    language already supports practically useful reasoning tasks.
\end{abstract}

%% file: section/introduction.tex
\section{Introduction}

Deriving new knowledge from existing data is known as reasoning or inference.
RDF reasoning has traditionally relied on fixed entailment regimes,
such as RDFS and OWL. While powerful for formal ontology modeling,
these regimes are rigidly tied to built-in W3C vocabularies, suffer from
high computational complexity~\cite{glimm2011using}, and operate under the
Open World Assumption, which prevents practical tasks like negation as
failure. Rule-based alternatives appeared early but without a shared
standard, adoption fragmented into engine-specific dialects. The proposed
SRL specification in the W3C Data Shapes Working
Group~\cite{ShaclRules2026} brings
user-defined production rules into the SHACL family. Several reasoners
predate it (surveyed in Section~\ref{sec:related-work}), but SRL is, to
our knowledge, the first W3C-backed specification for rule-based RDF
inference.

We approach the specification from the implementer's perspective, aiming
to give the reader a clear picture of the building blocks a rule reasoner
needs: parsing, dependency analysis, stratification, and fixpoint
evaluation.

To explore these mechanics we built two engines. The first reuses SPARQL
infrastructure by extending the Comunica query
engine~\cite{TaelmanComunica2020}. The second, Eyeleng~\cite{Eyeleng2026}, is a dedicated
engine written from scratch in JavaScript that automatically combines
forward materialization with backward goal-directed evaluation. Both were
validated against the SRL conformance test suite and a set of
use-case-specific tests. Live demos are available at
\url{https://comunica.github.io/comunica-feature-shacl-rules/shacl-ui/public/}
(Comunica-based) and \url{https://eyereasoner.github.io/eyeleng/playground}
(Eyeleng). While our implementations target SRL concretely, our insights
apply broadly to SPARQL-based rule engines. This paper
compares the architectural trade-offs and surfaces key lessons about
stratification, output handling, SPARQL compatibility, and whether SRL
paves the road to explainable reasoning.

The remainder of the paper is structured as follows.
Section~\ref{sec:related-work} surveys existing RDF reasoners.
Section~\ref{sec:shacl-rules} introduces the relevant SRL concepts.
Section~\ref{sec:implementations} describes the two engines.
Section~\ref{sec:design-choices} presents our findings and lessons
learned.

%% file: section/related_work.tex
\section{Related work}
\label{sec:related-work}

Reasoning engines in the Semantic Web are typically deployed to infer new implicit knowledge and map between vocabularies.
Yet, while reasoning is core to these use cases, the community has never settled on a single language.
Instead, three distinct paradigms of standardization have evolved in parallel:

\begin{description}
    \item[Entailment regimes] define fixed inference rules over
    vocabularies. RDFS entailment derives class membership and
    property hierarchies. OWL 2 RL~\cite{Owl2Profiles2009} extends
    this with a Datalog-expressible profile of OWL designed for
    rule-based implementations. These are \emph{meta-vocabularies}:
    they specify what follows from RDFS or OWL axioms, not how to
    write custom rules. SPARQL itself officially supports querying
    under such entailment regimes~\cite{glimm2011using}, but this
    only exposes the fixed RDFS/OWL semantics at query time---it does
    not let users define their own inference rules.

    \item[Rule languages] give users a syntax for writing their own
    implications. Datalog~\cite{Datalog1989} is the theoretical
    foundation---Horn clauses over relational atoms with guaranteed
    termination. N3~\cite{berners2008n3logic} extends RDF with
    quoted graphs and \texttt{=>} implications, supported by engines
    such as EYE and CWM. SWRL~\cite{Swrl2004} combines OWL with
    Horn clauses but requires an OWL reasoner. RIF~\cite{RifCore2013}
    is a W3C standard for rule interchange spanning multiple
    dialects, though its complexity limited adoption.

    \item[SPARQL-based rules] embed rules directly in SPARQL syntax,
    making them accessible to existing SPARQL engines. SPIN~\cite{Spin2011}
    represented CONSTRUCT queries as RDF triples linked to classes.
    SRL builds on this lineage: it adds stratification,
    negation-as-failure, and a stand-alone textual syntax, but
    inherits SPARQL's expression language and evaluation semantics.
\end{description}

%% file: section/shacl_rules.tex
\section{SRL}
\label{sec:shacl-rules}

SRL defines production rules that derive new RDF triples from a given
base graph. A rule consists of a body (a graph pattern to match against the
data) and a head (triple templates whose variables are filled in by matching
solutions). Rules are organized into rule sets, which may also import other
rule sets and include inline data blocks. The spec provides two concrete
syntaxes: a human-readable Shape Rules Language (SRL/text) and an RDF
representation (SRL/RDF). This paper works with the 11 July 2026 Working Draft of the SRL
spec~\cite{ShaclRules2026}. A minimal rule set in SRL/text looks as follows:

\begin{lstlisting}
PREFIX : <http://example.org/>

DATA { :alice :parentOf :bob . }

RULE { ?child :childOf ?parent }
WHERE { ?parent :parentOf ?child }
\end{lstlisting}

\subsection{Foundational Concepts}

The spec defines two operations---\emph{infer} (materialize the full
inference graph) and \emph{query} (check whether a goal pattern is
derivable)---but only infer receives a defined evaluation procedure.

Evaluation proceeds to a fixpoint: rules fire repeatedly until no
new triples emerge, with each conclusion immediately visible to
subsequent rules. To prevent non-determinism---in particular, to ensure that
negation-as-failure and assignment produce the same result regardless of rule
execution order---the spec requires \emph{stratification}. A dependency graph
is built over the rules, with edges labeled \emph{open} (the caller can
handle incremental additions from the callee) or \emph{closed} (the callee
must fully complete before the caller runs).

For example, if Rule~A produces \texttt{:childOf} triples and Rule~B reads
\texttt{:childOf} in a regular body pattern, the dependency is open: B may
miss some matches on a given iteration, but the fixpoint loop lets it catch
up once A produces more data---missing data is temporary, not incorrect.
If Rule~B instead tests \texttt{NOT \{ ?x :childOf ?y \}}, the dependency
becomes closed: a premature absence would cause B to draw a conclusion that
the fixpoint cannot later retract, so A must fully complete before B begins.

Stratification partitions the rule set into an ordered sequence of layers
(strata). Each layer contains \emph{run-once} rules (those with assignments
or blank-node heads) followed by \emph{general} rules evaluated to a fixpoint
within that layer. Lower layers complete before higher layers begin,
guaranteeing that negation blocks, assignment expressions, and blank-node
creation see only fully-computed data from earlier strata. Rule sets that
violate the stratification condition (a closed edge in a dependency cycle)
are ill-formed and have no defined outcome.

The SRL body language is a restricted subset of SPARQL: triple patterns,
FILTER, NOT, and SET, but no OPTIONAL, no property-path \texttt{*} and
\texttt{+}, and a reduced built-in function set. This deliberate overlap
means an SRL engine can be built largely by reusing SPARQL
infrastructure.

\subsection{Why the Data Shapes Working Group Addressed Rules}

The Data Shapes Working Group was re-chartered in December 2024 to extend
SHACL with standards-track reasoning
capabilities~\cite{DataShapesCharter2024}. The charter lists ``reasoning
rules, their dependencies, ordering and packaging'' as within scope. The
motivation is straightforward: a SHACL shape already encodes a graph
pattern, making it natural to repurpose that pattern as the body of an
inference rule.

%% file: section/implementations.tex
\section{Implementations}
\label{sec:implementations}

We discuss two implementations: one reusing SPARQL infrastructure inside
the Comunica query engine, and one built from scratch in JavaScript without
any SPARQL engine underneath. The goal in both cases is architectural: a
reader does not need to follow every implementation detail, only the shape
of the pipeline, so that the same building blocks can be re-assembled with
different algorithms.

\subsection{Comunica-based engine — pipeline and algorithms}

\begin{figure}[ht]
  \centering
  \includegraphics[width=\textwidth]{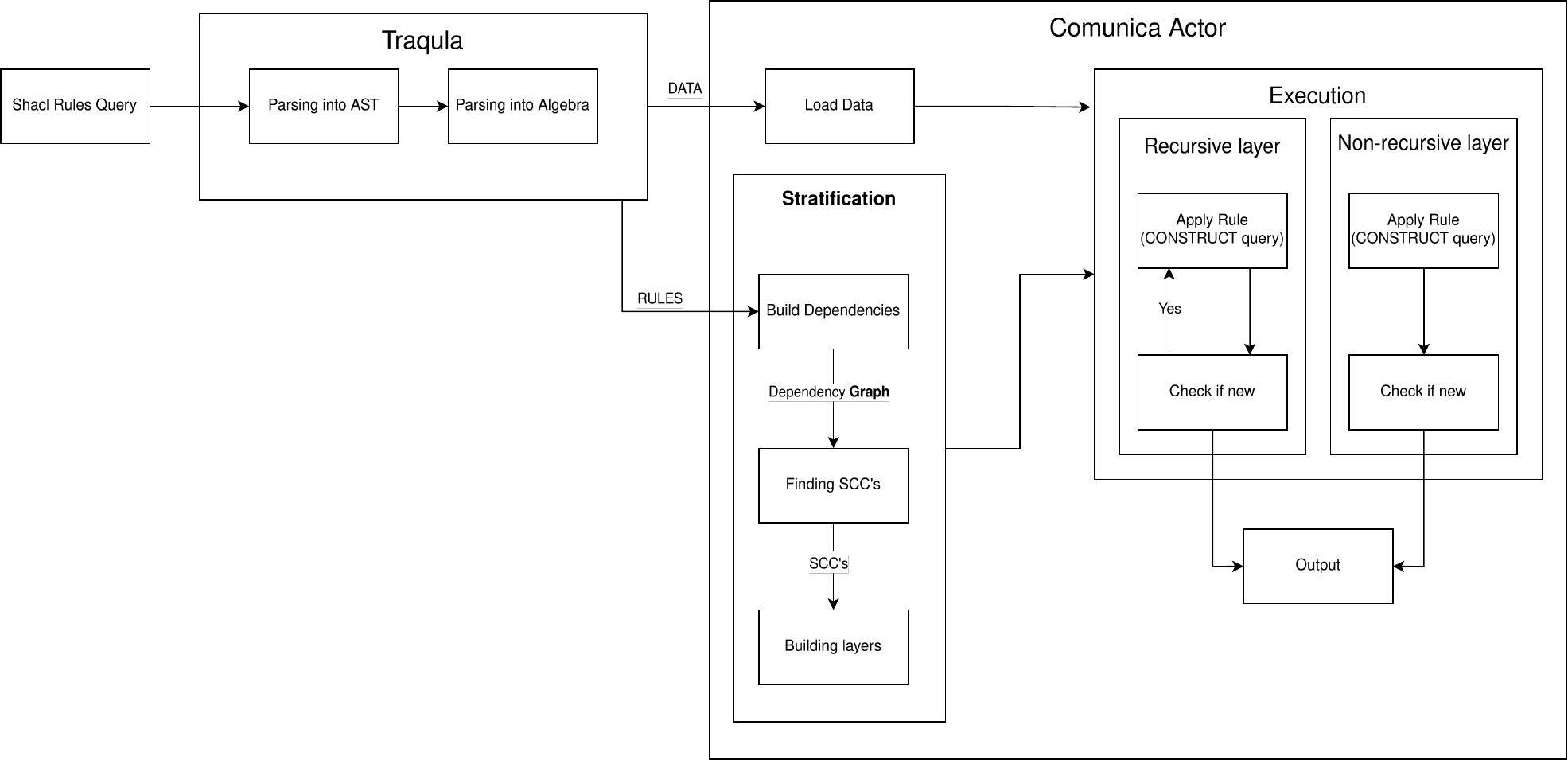}
  \caption{Execution pipeline: SRL document $\rightarrow$ parse actor $\rightarrow$
    algebra $\rightarrow$ dependency analysis (inverted index + Kosaraju SCC +
    Kahn topological sort) $\rightarrow$ stratified layers $\rightarrow$ CONSTRUCT fixpoint
    $\rightarrow$ inferred quads.}
  \label{fig:pipeline}
\end{figure}

We built an SRL engine as an extension of Comunica, a modular
SPARQL query engine. Comunica's actor-bus architecture
allowed us to add two components without modifying its core: a
\emph{parse actor} that converts SRL text into SPARQL algebra, and
an \emph{operation actor} that executes the rule set via
forward-chaining CONSTRUCT operations.

\subsubsection{Pipeline}

Figure~\ref{fig:pipeline} shows the execution pipeline. A SRL
document enters the parse actor, which uses Traqula \cite{de2026traqula}
to produce a SPARQL algebra tree. The algebra node has type
\texttt{shaclRule} and contains a \texttt{data} array (ground
triples from DATA blocks) and a \texttt{rules} array (each rule
with a body algebra and head templates).

After loading DATA-block triples into the in-memory store, the
operation actor executes the algebra in three phases:

\begin{enumerate}
\item \textbf{Analyse dependencies}. The engine walks every rule's
  body algebra to extract triple patterns, then checks which
  rules' head templates could produce triples matching those
  patterns. This produces a directed dependency graph where an
  edge $A \rightarrow B$ means rule~$A$ can consume output from
  rule~$B$, so $B$ must execute before $A$.
\item \textbf{Stratify}. The dependency graph is partitioned into
  evaluation layers. Kosaraju's SCC algorithm finds
  strongly-connected components in $O(V+E)$, then Kahn's
  topological sort orders them into layers in $O(V+E)$. This
  separates ``which rules form a mutual-recursion cluster'' from
  ``in which order do the clusters fire,'' reaching the same
  stratification layers as the spec's iterative algorithm
  (\S4.4.2) while being easier to debug and profile. The spec
  permits this substitution explicitly: conformance ``depends on
  producing a dependency graph that meets the definitions of a
  dependency graph, not on the use of this procedure'' (\S4.3.2). Layers
  containing intra-component cycles are flagged as recursive.
\item \textbf{Execute layers}. Non-recursive layers run each rule once
  in topological order; recursive layers fixpoint-loop until no new
  quads are produced. Every rule fires as a SPARQL CONSTRUCT
  operation, with newly inferred quads inserted immediately so
  subsequent rules can consume them.
\end{enumerate}

\subsubsection{Dependency detection}

The dependency analysis at the heart of phase~2 checks whether a
head template can possibly generate a triple that matches a body
pattern. This is the spec's ``triple pattern matching'' rule
(\S4.3): a template component (subject, predicate, or object)
matches a pattern component if either is a variable---which acts
as a universal acceptor---or both are identical concrete RDF
terms. Let $R$ be the number of rules. A brute-force
implementation checks every pair of rules, at $O(R^2)$ cost.

To make this practical, we build an inverted index of head
templates. Three hash maps store rule IDs keyed by the concrete terms they
produce in each position, plus three catch-all sets listing rules that use
a variable in that position:

\begin{lstlisting}[language=Java]
// Rule A:  { ?x :childOf ?y }  WHERE { ?x :executes ?y }
// Rule B:  { ?x :trusts   ?y }  WHERE { ?x :childOf  ?y }
\end{lstlisting}

Given these rules, the predicate-index maps \texttt{:childOf} to
\{A\} and \texttt{:trusts} to \{B\}. When we process a body pattern---say,
\texttt{?x :childOf ?y} from Rule~B's body---we check each position: the
predicate \texttt{:childOf} maps to one rule (A), while the subject and
object are variables and would match all $R$ rules. We pick the most
restrictive position (predicate, one candidate) and only check that rule.

With $n$ the maximum number of candidate rules returned by the most
selective position of any body pattern, the total cost drops from
$O(R^2)$ to $O(R \times n)$. In our deep-taxonomy use case tests, rules
form a linear dependency chain where each rule's body depends on at most
one head template, making $n = 1$: with 100,000 rules, the dependency
analysis shrinks from 10~billion pairwise checks to 100,000.\footnote{Try
the deep-taxonomy test at \url{https://comunica.github.io/comunica-feature-shacl-rules/shacl-ui/public/\#/examples-tests}.}

\subsection{Eyeleng — a from-scratch hybrid engine}

The second implementation is Eyeleng~\cite{Eyeleng2026} (Logic Engine Next
Generation), a standalone SRL engine written in
JavaScript without any SPARQL engine underneath. Unlike the Comunica
approach, which delegates all evaluation to an existing SPARQL pipeline,
Eyeleng implements parsing, dependency analysis, stratification, and rule
execution from scratch, giving it full control over the evaluation
strategy.

The key architectural difference is \emph{hybrid execution}. Comunica fires
every rule as a forward CONSTRUCT query against the accumulating store—pure
bottom-up materialization. Eyeleng instead analyzes the dependency graph
and decides per-rule: ordinary consequences are materialized forward, while
predicates that act as function-like computations (rules containing
assignments whose output feeds other rules) are evaluated backward,
on-demand, using tabling. This avoids publishing intermediate helper triples
that were only needed during computation---the engine keeps them internal and
outputs only the final inferred graph.

\subsection{Comparison}

The two engines embody opposite trade-offs. Comunica inherits everything
from its SPARQL host: federated queries, heterogeneous input formats
(Turtle, JSON-LD, TriG), browser bundling via Webpack, and automatic
improvements whenever the underlying SPARQL engine gains faster joins or
better indexing. The cost is that every rule body must travel through the
full SPARQL query pipeline---parsing a CONSTRUCT, building algebra, and
planning execution---on every firing. Backward reasoning is theoretically
possible thanks to the actor-bus architecture but was not pursued, as one
goal was to demonstrate how far pure SPARQL reuse reaches. The output is a
flat quad stream with no built-in provenance.

Eyeleng takes the opposite approach: it evaluates rules directly against
its own store, skipping the SPARQL pipeline and running several times faster
(Section~\ref{sec:performance}). Its hybrid forward/backward execution keeps intermediate
computations internal and produces proof trees as a byproduct of backward
resolution (discussed in Section~\ref{sec:design-choices}). The trade-off
is that every piece of infrastructure---federated queries, input format
support, browser bundling---must be implemented from scratch.

Both engines target the same spec and are validated against the same
conformance suite (Section~\ref{sec:design-choices}). Neither approach is
prescribed by the spec; both surface different questions about what an SRL
engine should provide beyond correctness.

\subsection{Performance evaluation}
\label{sec:performance}

\subsubsection{Setup and protocol}

Both engines were measured on the same machine: an AMD Ryzen~5 5500U
(6~cores, 12~threads), 15~GiB RAM, Linux~6.8.0, Node.js~v20.20.2 (V8~11.3),
running \texttt{@comunica/query-shacl-rule}~1.0.0 with
\texttt{rdf-stores}~2.2.0 and Eyeleng~1.2.2.

Both are timed in-process over identical boundaries: the clock starts with the
rule set in memory as a string and stops once the full closure has been
materialised into an array. Both therefore include parsing, dependency
analysis, stratification, fixpoint evaluation and result materialisation, and
neither includes process start-up, file reading or serialisation. Each engine
receives its own discarded warm-up run, and engine order alternates between
workloads.

Each workload ran 25 times, except the six largest (5--22 runs); we report
medians with the interquartile range. Cost models are fitted to the
per-workload minima, which are less distorted by garbage-collection pauses.
Each suite varies one factor and holds the others fixed, and both engines
produced identical output triple counts on every workload. The harness, the generated rule sets and the raw timings are
available in the
repository.\footnote{\url{https://github.com/comunica/comunica-feature-shacl-rules}, directory \texttt{performance/}.}

\subsubsection{Results}

Table~\ref{tab:perf} reports all 25 workloads. Eyeleng is faster on every one,
by 2.15$\times$ to 6.44$\times$; there is no crossover point.

\begin{table}[ht]
\centering
\small
\caption{Median execution time in ms, interquartile range in parentheses.
Speedup is Comunica's median divided by Eyeleng's. 25~timed runs after one
discarded warm-up, except the six largest workloads (5--22 runs).}
\label{tab:perf}
\begin{tabular}{l r@{\ }l r@{\ }l r}
\hline
Workload & \multicolumn{2}{c}{Comunica} & \multicolumn{2}{c}{Eyeleng} & Speedup \\
\hline
\multicolumn{6}{l}{\emph{Rule-set size} --- deep-taxonomy chains} \\
\quad 10 rules      &     15.2 & (3.0)   &     2.6 & (1.2)   & 5.79 \\
\quad 100 rules     &     43.1 & (11.4)  &     7.4 & (5.4)   & 5.83 \\
\quad 1,000 rules   &    274.3 & (17.7)  &    47.1 & (11.6)  & 5.83 \\
\quad 10,000 rules  &  2,838   & (55)    &   546.9 & (89.9)  & 5.19 \\
\quad 100,000 rules & 31,962   & (578)   & 6,279   & (232)   & 5.09 \\
\hline
\multicolumn{6}{l}{\emph{Data-graph size} --- 2 rules, one fixpoint pass} \\
\quad 1K triples    &     67.8 & (4.7)   &    16.0 & (6.0)   & 4.24 \\
\quad 10K triples   &    696.9 & (23.0)  &   246.7 & (46.8)  & 2.83 \\
\quad 100K triples  &  7,862   & (103)   & 2,390   & (251)   & 3.29 \\
\hline
\multicolumn{6}{l}{\emph{Recursion depth} --- transitive closure over a chain} \\
\quad 25 nodes      &     65.6 & (4.5)   &    30.5 & (6.4)   & 2.15 \\
\quad 50 nodes      &    560.9 & (12.4)  &   204.2 & (3.0)   & 2.75 \\
\quad 100 nodes     &  4,511   & (47)    & 1,623   & (11)    & 2.78 \\
\quad 150 nodes     & 14,652   & (42)    & 5,774   & (53)    & 2.54 \\
\quad 200 nodes     & 34,510   & (34)    & 14,135  & (421)   & 2.44 \\
\hline
\multicolumn{6}{l}{\emph{Body complexity} --- 200 rules, 10K in, 2K out, fixed} \\
\quad 1 pattern     &    564.7 & (37.4)  &   112.2 & (47.5)  & 5.03 \\
\quad 3 patterns    &    678.8 & (69.3)  &   166.0 & (73.1)  & 4.09 \\
\quad 5 patterns    &    846.4 & (107.5) &   170.9 & (36.0)  & 4.95 \\
\hline
\multicolumn{6}{l}{\emph{Firing count} --- 6K in, 6K out, 1 pattern/body, fixed} \\
\quad 10 rules      &    374.7 & (26.4)  &    94.4 & (19.8)  & 3.97 \\
\quad 50 rules      &    379.4 & (21.6)  &    90.4 & (19.7)  & 4.19 \\
\quad 200 rules     &    429.7 & (30.8)  &    97.0 & (30.4)  & 4.43 \\
\quad 500 rules     &    456.3 & (24.8)  &   102.1 & (23.8)  & 4.47 \\
\quad 1,000 rules   &    547.4 & (31.7)  &   112.3 & (18.4)  & 4.88 \\
\quad 1,500 rules   &    656.2 & (28.3)  &   116.1 & (10.2)  & 5.65 \\
\quad 2,000 rules   &    770.8 & (72.7)  &   131.8 & (36.8)  & 5.85 \\
\quad 3,000 rules   &    967.4 & (52.6)  &   150.2 & (17.0)  & 6.44 \\
\quad 6,000 rules   &  1,541   & (42)    &   240.5 & (78.8)  & 6.41 \\
\hline
\end{tabular}
\end{table}

\paragraph{Rule-set size.}
Across four orders of magnitude the speedup is essentially constant, drifting
from 5.8$\times$ at 10 rules to 5.1$\times$ at 100,000. Deep-taxonomy varies
rule count, data and derivation depth together, so it shows that the gap is
stable, not what causes it. The next two suites isolate the cause.

\paragraph{Cost per rule firing.}
The firing-count suite splits the same 6,000 matches across 10 to 6,000 rules,
holding input size, output size and body width identical; only the number of
firings changes. Both engines are linear across the full range, with very
different slopes: Comunica pays \textbf{193~$\mu$s per rule firing}
($R^2 = 0.999$), Eyeleng \textbf{20~$\mu$s} ($R^2 = 0.995$)---a factor of 9.6.
Routing every rule body through the full SPARQL pipeline, parsing a CONSTRUCT,
building algebra and planning execution, imposes a fixed cost per firing that a
direct evaluator avoids.

Because both are linear, the speedup is a ratio of two straight lines: it rises
from 4.6$\times$ at 10 rules to 7.6$\times$ at 6,000 and converges towards the
slope ratio of 9.6$\times$ rather than growing without bound. A single speedup
figure for these engines is therefore meaningless without the rule count it was
measured at. Two caveats: the fitted intercepts differ by 4.6$\times$ (344~ms
against 75~ms), so per-firing overhead is not the whole gap; and holding total
work fixed while raising rule count drops the matches per rule from 600 to 1,
so the top of the range does not represent rule sets in which each rule does
substantial work.

\paragraph{Body complexity.}
Holding rule count (200), input (10,000 triples) and output (2,000 triples)
constant while varying only the patterns per body isolates the same effect on a
second axis. Comunica pays \textbf{263~$\mu$s per pattern per rule}
($R^2 = 1.000$) against Eyeleng's \textbf{50~$\mu$s} ($R^2 = 0.998$), a factor
of 5.3. Comunica's per-pattern and per-firing costs are of the same order,
consistent with a pipeline cost that scales with the size of the algebra tree
rather than with the number of solutions it produces.

\paragraph{Data-graph size.}
Both engines scale linearly with the size of the input graph---doubling the
data doubles the runtime---and the gap is narrowest here:
2.8--4.2$\times$, against 4--6.4$\times$ in the rule-bound suites. At 100K
triples Comunica spends 79~$\mu$s per input triple and Eyeleng 24~$\mu$s. The
cost of reusing SPARQL infrastructure is paid per rule and per pattern, not per
triple: it amortises on data-heavy workloads and bites hardest on large rule
sets.

\paragraph{Recursion depth.}
Runtime grows with the cube of the chain length in both engines, measured over
five chain lengths, because both re-evaluate every rule against the whole
accumulated store on each iteration rather than only against newly derived
triples. This suite has the smallest and
flattest gap (2.15--2.78$\times$): when an asymptotically suboptimal strategy
dominates, the architectural difference is largely irrelevant. The
specification mandates fixpoint semantics but is silent on how the fixpoint is
reached, and both implementations independently converged on the naive
strategy; semi-naive evaluation, standard in Datalog engines, would change
these numbers for both engines far more than their architecture does.

%% file: section/findings.tex
\section{Findings}
\label{sec:design-choices}

Eyeleng passes all 166 conformance tests. The Comunica-based implementation
passes 154: it computes the correct result for every test that asks it to
derive triples, and all twelve remaining cases are negative tests, which supply
input that a conformant engine is required to reject. Four are malformed rule
sets that our parser accepts rather than rejecting, and eight test
well-formedness and stratification conditions that we do not yet check. The
distinction matters for a paper about implementing the specification: the gap
is in input validation, not in evaluation, and some of these cases exposed
ambiguities in the grammar or issues in the test suite itself rather than
shortcomings of our engine.

The findings below are a mix of general lessons about the spec and observations
specific to building on top of an existing SPARQL engine; each subsection
notes which is which, and, where a finding is implementation-specific, how
it could be avoided in a different design.

\subsection{Explainability, Proof Trees, and the Under-defined Query Operation}

\emph{This is a spec-level finding: it applies to any SRL engine, not only to
ours.} The \texttt{query} operation, described in \S\ref{sec:shacl-rules}, is
named by the spec but never given an evaluation procedure.

This omission matters for explainability. Bottom-up materialization
(\texttt{infer}) carries no provenance by default: if a user asks ``Why is
\texttt{:X :descendedFrom :C}?'', the engine can only point at the output
stream, not the chain of rule firings behind it. This is not a hard
limitation---an implementation can instrument its CONSTRUCT evaluations to
tag every inferred triple with rule ID, iteration, and bindings---but it
requires explicit tracking that the spec does not mandate and that most
forward-only engines skip.

A top-down, goal-directed \texttt{query} operation would naturally produce
exactly that chain. In Datalog, evaluating \texttt{?- ancestor(xerces, X)} from
a set of rules and facts builds a proof tree: a step-by-step derivation showing
how each goal reduces to subgoals until facts are reached. For SRL, a
\texttt{query} operation could work the same way---given a goal pattern, prove
it by backward-chaining through rule bodies and data, yielding a tree of rule
applications and bindings. That tree \emph{is} the explanation.

Eyeleng supports explainability because its hybrid forward/backward
architecture tracks rule-to-output provenance. Comunica's forward-only
pipeline, which fires each rule as a standalone CONSTRUCT query against the
accumulating store, produces a flat stream of output quads with no recording of
derivation history. Adding provenance would require instrumenting every
CONSTRUCT evaluation, which is feasible but imposes overhead that is currently
outside Comunica's core design.

The key finding for the spec: if \texttt{query} is to serve as an
explainability primitive, the specification should define its output form---a
proof tree, a set of rule-to-triple bindings, or at minimum a provenance
model---rather than leaving it as an undefined term. Without this, every
implementer provides a different (or no) explanation mechanism, and
interoperability of explanations becomes impossible.

\subsection{Blank node syntax in WHERE clauses}
\label{sec:collections}

This finding is specific to the Comunica-based implementation, not a defect
of SRL itself: it arises from a mismatch between two independently
developed components (Traqula and Comunica's algebra evaluator), not from
any ambiguity in the spec.

Turtle's blank node property list syntax \texttt{[ ... ]} is a common
shorthand in both DATA blocks and WHERE bodies. Under the hood it
introduces a fresh blank node and attaches triples to it. The problem
surfaces when the same shorthand appears on both sides of a rule.
Handling blank nodes correctly is a known tension in the RDF
stack~\cite{HOGAN201442}: blank nodes are existential variables without
fixed identifiers, yet every system must assign them concrete labels to
represent them. Our mismatch---Traqula assigning fixed labels to algebra
patterns while Comunica treats them as identity checks---is the same
tension playing out at the component boundary.

\begin{lstlisting}
DATA   { :alice :address [ :city "Paris" ] . }
RULE   { :alice :city ?city }
WHERE  { :alice :address [ :city ?city ] }
\end{lstlisting}

The DATA block expands to two triples with a blank node---say,
\verb|_:b1 :city "Paris"| and \verb|:alice :address _:b1|---which are
inserted into the store. Traqula expands the WHERE body to the same
structure, but assigns a different blank node label---say,
\verb|_:g1|---during algebra construction. Comunica's algebra-based BGP
matching treats blank nodes as concrete identifiers: \verb|_:g1| does
not equal \verb|_:b1|, so no triple in the store satisfies the pattern
and the rule produces zero output.

Replacing the WHERE shorthand with a variable solves it:

\begin{lstlisting}
WHERE { :alice :address ?addr . ?addr :city ?city }
\end{lstlisting}

The same failure mode occurs with Turtle collection syntax \texttt{( ... )},
since it is also syntactic sugar over blank nodes.

Eyeleng avoids both problems entirely: its parser always expands
property-list and collection syntax to variable-based triple patterns,
so no blank node labels cross the parser-to-engine boundary. Traqula, as
an external algebra builder, produces blank node RDF terms that
Comunica's algebra path cannot distinguish from concrete references.
Fixing this requires either making Comunica's algebra BGP path treat
blank nodes as existential variables (as its SPARQL path already does),
or making Traqula produce variable-based patterns for both shorthand
forms.

%% file: section/conclusion.tex
\section{Conclusion}

We set out to answer a simple question: how much of existing SPARQL
infrastructure can be reused to build a rule engine? The answer is most
of it, but the remaining gap is exactly where the interesting problems
live.

Building on Comunica gave us federated queries, heterogeneous input
formats, browser bundling, and a battle-tested SPARQL expression evaluator
for free---at the cost of explainability and output cleanliness. Eyeleng gets those
capabilities for free by being purpose-built, reimplementing every piece of
infrastructure from scratch, and runs faster as a result of skipping the
SPARQL query pipeline on every rule firing.

Two findings transcend the choice of engine. The spec's \texttt{query}
operation is named but never defined; giving it goal-directed semantics
with a proof-tree output would standardise explainability across
implementations. The open/closed distinction in the dependency graph
determines whether rules share a stratum or must be split, and our
Kosaraju+Kahn decomposition shows the spec's iterative algorithm can be
replaced without changing semantics. On the implementation side, blank
node syntax in WHERE bodies exposes a mismatch between how an external
algebra builder produces patterns and how Comunica's algebra path
consumes them---a reminder that reusing SPARQL infrastructure means
inheriting its edge cases along with its strengths.

SRL has the bones of a practical rule language: stratification,
fixpoint semantics, and dependency analysis are handled well. But what
provenance looks like and how \texttt{query} works are questions
implementers must answer for themselves today.